# AI-Native Industrial Control Systems: An Agent Architecture for the Control System Lifecycle

Lisi Liu, Ang Zhou, Chuan Pang, Feng Gao, and Shengkai Liu

Department of New Networks, Pengcheng Laboratory, Shenzhen, China

**Abstract**

Industrial control systems increasingly use AI for perception, control, and engineering, while available control functions and execution environments are typically established in advance. This paper proposes an AI-native industrial control system in which collaborating agents determine the capabilities required by a task, construct the control application, and prepare its execution environment. System-level agents coordinate device-level agents within configured resource and permission constraints. A physical sorting workstation illustrates two forms of adaptation. In the first case, controller and robot agents construct a sorting application on a prescribed Raspberry Pi target. In the second, a camera agent adds inspection on an edge node, and the controller associates inspection results with downstream pickup decisions. A subsequent revision enables selective sorting using the same agents and deployment arrangement. The trials demonstrate the feasibility of task-driven construction and evolution on the workstation. Commissioning also reveals limitations in the traceability of design decisions, adherence to delegated responsibilities, diagnostic reasoning, and verification of physical behavior.

## 1. Introduction

Artificial intelligence (AI) is becoming an important enabler of industrial transformation, supporting manufacturing activities from engineering and production to maintenance. At the control level, AI is increasingly being integrated into industrial control systems (ICSs), including systems based on programmable logic controllers (PLCs) and computer numerical controllers (CNCs). Existing studies can be grouped into three categories according to the primary role of AI, as summarized in Fig.1a):

- AI for perception and estimation. AI processes sensor or process data to provide information used by control systems, such as visual inspection results[1], estimated process states[2], and predicted tool conditions[3].
- AI for control. AI computes or adapts control actions, controller parameters, or control policies. Examples include fuzzy control[4], neural network control[5], and reinforcement learning-based control[6].
- AI for engineering and operation. AI supports control program generation and verification, commissioning, troubleshooting, and maintenance guidance. Recent studies increasingly employ large language models (LLMs) to assist with or automate these tasks, with the potential to reduce engineering effort and improve operational efficiency [7-9].

LLM-based agents further extend these functions toward iterative task execution by combining reasoning, tool use, and feedback. Recent studies use agents to interpret project context, generate and compile control code, check its behavior, and revise it in response to diagnostic results [10,11].

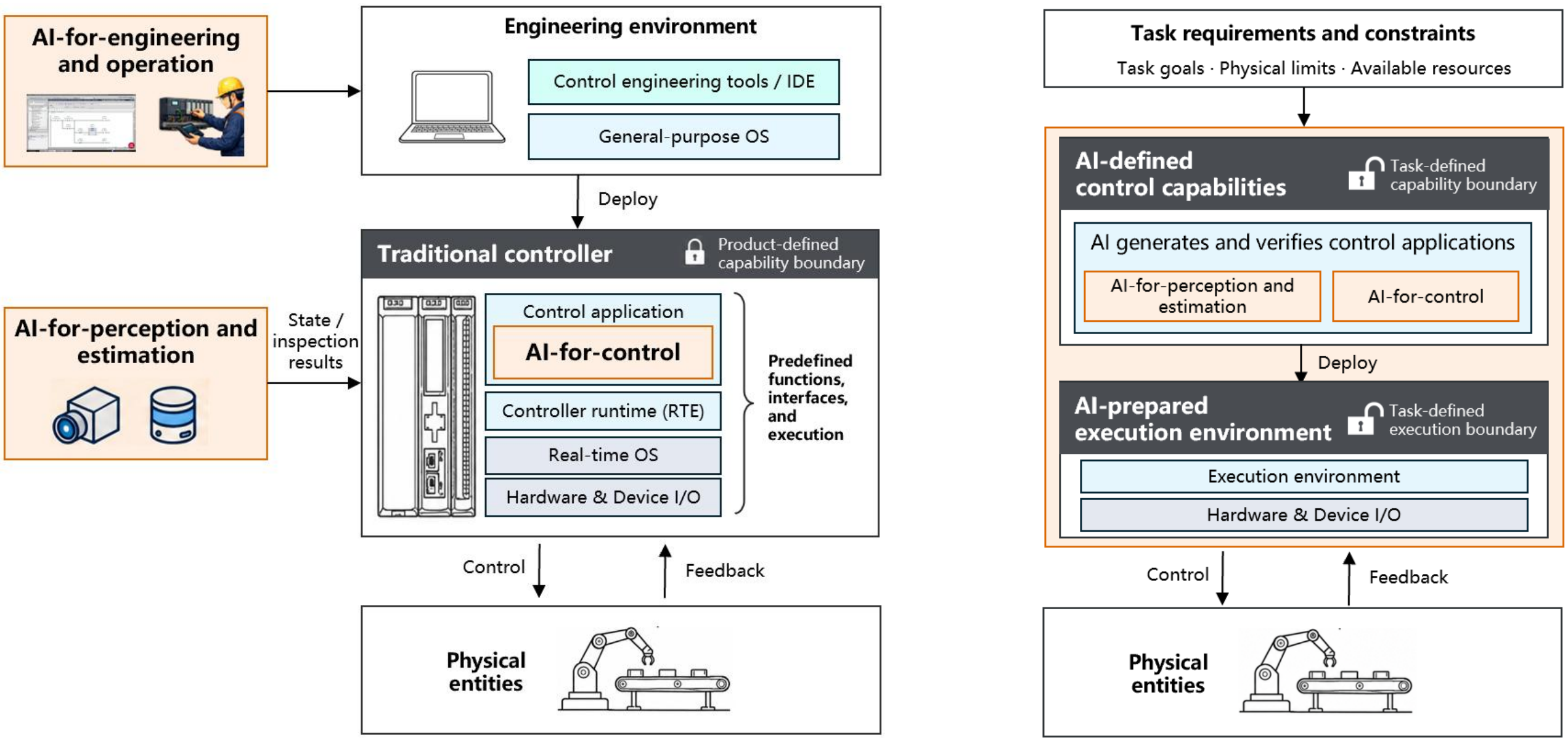

a) AI-augmented ICS: application development within product-defined boundaries

b) AI-native ICS: application development within task-defined capabilities and execution boundaries

Figure 1. Roles of AI in AI-augmented and AI-native industrial control systems.

Despite these advances, the integration pattern outlined above is primarily AI-augmented: **AI implements or optimizes control applications within control capability and execution boundaries established by the underlying ICS.** The capability boundary specifies the functions and interfaces available for application development, while the execution boundary constrains the runtimes, platforms, and deployment configurations that support those applications. AI may improve perception, implement control algorithms, or automate engineering tasks, but the control system's architecture and execution mechanisms remain largely determined through prior engineering. Consequently, advances in AI do not necessarily extend to determining how the control system itself is constructed, deployed, and evolved.

In this context, this paper proposes the concept of AI-native ICS, which extends AI's responsibility to defining control capability and execution boundaries, as well as constructing and deploying control applications, as illustrated in Fig.1b). Based on task requirements, physical constraints, and available infrastructure resources, AI serves the entire lifecycle of control systems. It determines the required capabilities and generates the verified control applications. It also inspects and prepares execution environments and supports control applications' deployment and monitoring. When requirements or operating conditions change, AI revises the capabilities, applications, or deployment arrangements accordingly.

Compared to AI-augmented, the proposed AI-native ICS therefore introduces flexibility in three related aspects:

a. Task-defined control capability boundary. Control capabilities can be adapted according to task requirements, rather than being confined to the functions predefined by a controller product.
b. Adaptable execution environment boundary. Execution environments are inspected and prepared on demand within the configured infrastructure resources and permission constraints. Where multiple deployment targets are available, AI can assess their suitability and select a deployment arrangement according to connectivity, compatibility, and execution requirements. Application placement is therefore not necessarily fixed to a predefined controller platform.
c. Coordinated evolution of capabilities and execution environments. Changes in control functionality and deployment can be considered together. AI can revise control logic, adapt interfaces, and reconfigure execution environments to meet updated task objectives, with affected implementations revalidated before deployment.

These flexible boundaries remain subject to physical constraints, resource availability, and execution requirements. AI-native

ICS does not require non-deterministic model reasoning to govern each physical action. Verified control applications execute on target platforms under defined control semantics and timing constraints. AI may still provide perception, estimation, or control functions within the application, while explicit acceptance conditions, response deadlines, and fallback behavior govern how its outputs affect execution. **The distinction lies in extending AI's role from operating within an established control system to participating in its lifecycle, from construction to evolution.**

This paper contributes an architecture for task-driven control system construction, a collaboration model for system-level and device-level agents, and a physical demonstration of construction and evolution. Section 2 overviews the proposed AI-native ICS, including the architecture, agent organization and skills, and control system construction and evolution process. Section 3 evaluates them on a sorting workstation based on a defined evaluation criterion. Section 4 relates the findings to existing architectures and examines implementation issues and research challenges. Section 5 concludes the paper and outlines future work.

# 2. Overview of AI-native ICS

## 2.1 Architecture of the AI-native ICS

As illustrated in Fig.2, the proposed AI-native ICS comprises a network of agentic AIs and an AI-native infrastructure that supports their execution. Agents, such as agentic controller, camera, and robot, may be supplied by different providers. Through collaboration, they determine the control capability and execution environment boundaries according to task requirements, physical constraints, and available resources.

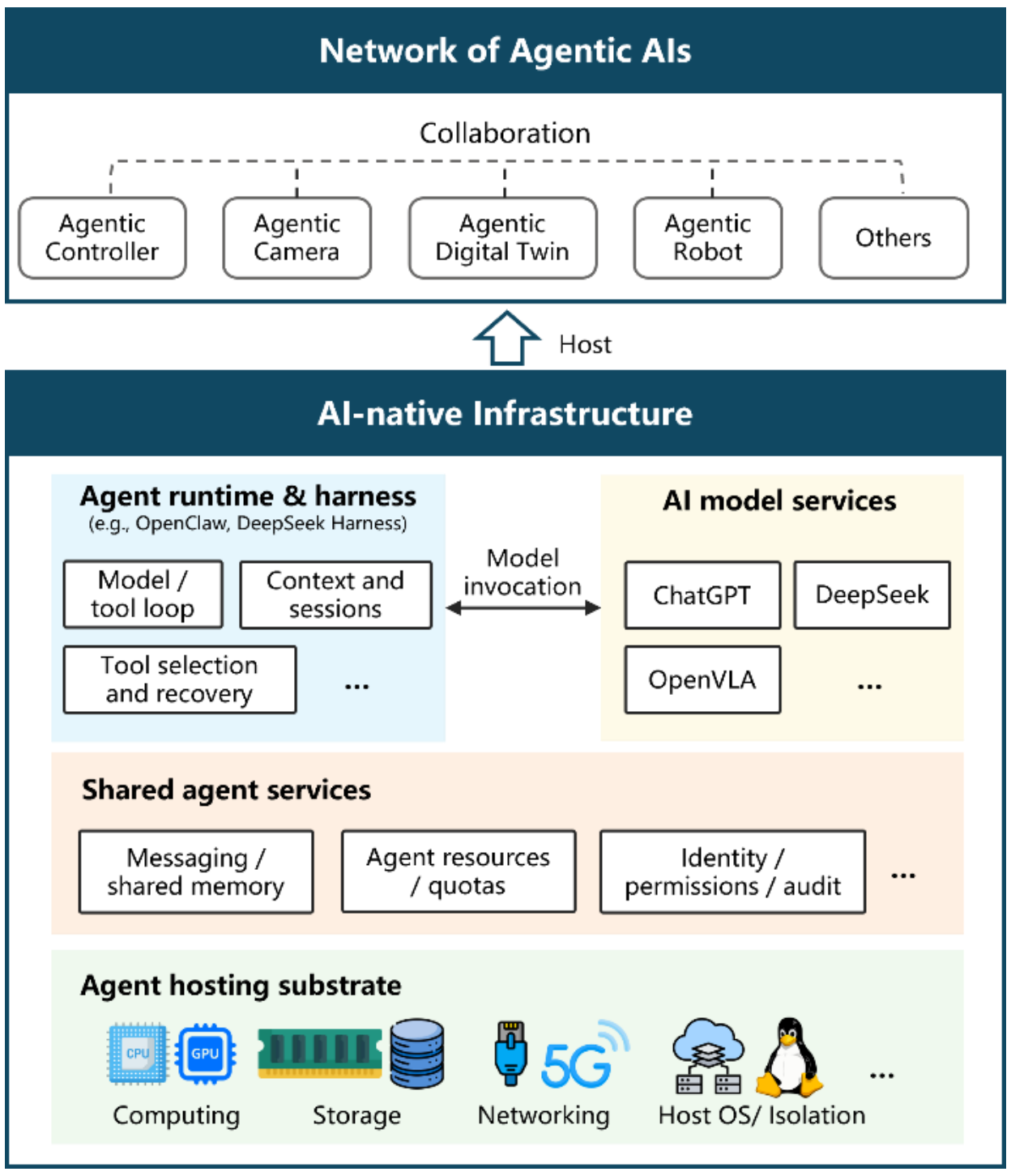

Figure.2 Architecture of the AI-native ICS

The AI-native infrastructure comprises four logical components. The agent runtime and harness manage model-tool execution loops, context and sessions, tool invocation, and recovery. AI model services provide local or remote inference capabilities. Shared agent services support messaging, shared memory, agent resource allocation and quotas, identity, permissions, and auditing. The agent hosting substrate supplies computing, storage, networking, and host operating-system and isolation mechanisms. These components represent logical responsibilities, that means a particular implementation may integrate several of them.

The AI-native infrastructure hosts the agents. The target infrastructure executes the applications they construct. These roles may share physical resources. The configuration can prescribe a deployment target or identify eligible resources from which the agents select a target according to connectivity, compatibility, and execution requirements.

## 2.2 Organization and Skills of Agentic AIs

Agentic AIs are classified as system-level or device-level according to their scope of responsibility. System-level agents address objectives spanning multiple devices or subsystems, including task decomposition, coordination, and capability composition. Device-level agents address device-specific capabilities and constraints, including control logic generation, target-platform adaptation, and operational management. For instance, Agentic Robot and Agentic Camera are device-level agents while Agentic Controller is a system-level agent.

As illustrated in Fig.3a), the agent network is dynamically organized according to task requirements. Both system-to-system and system-to-device associations are established from the assigned roles, coordination scopes, and permissions, rather than being fixed by a predefined topology. These associations can be modified when the governing configuration changes.

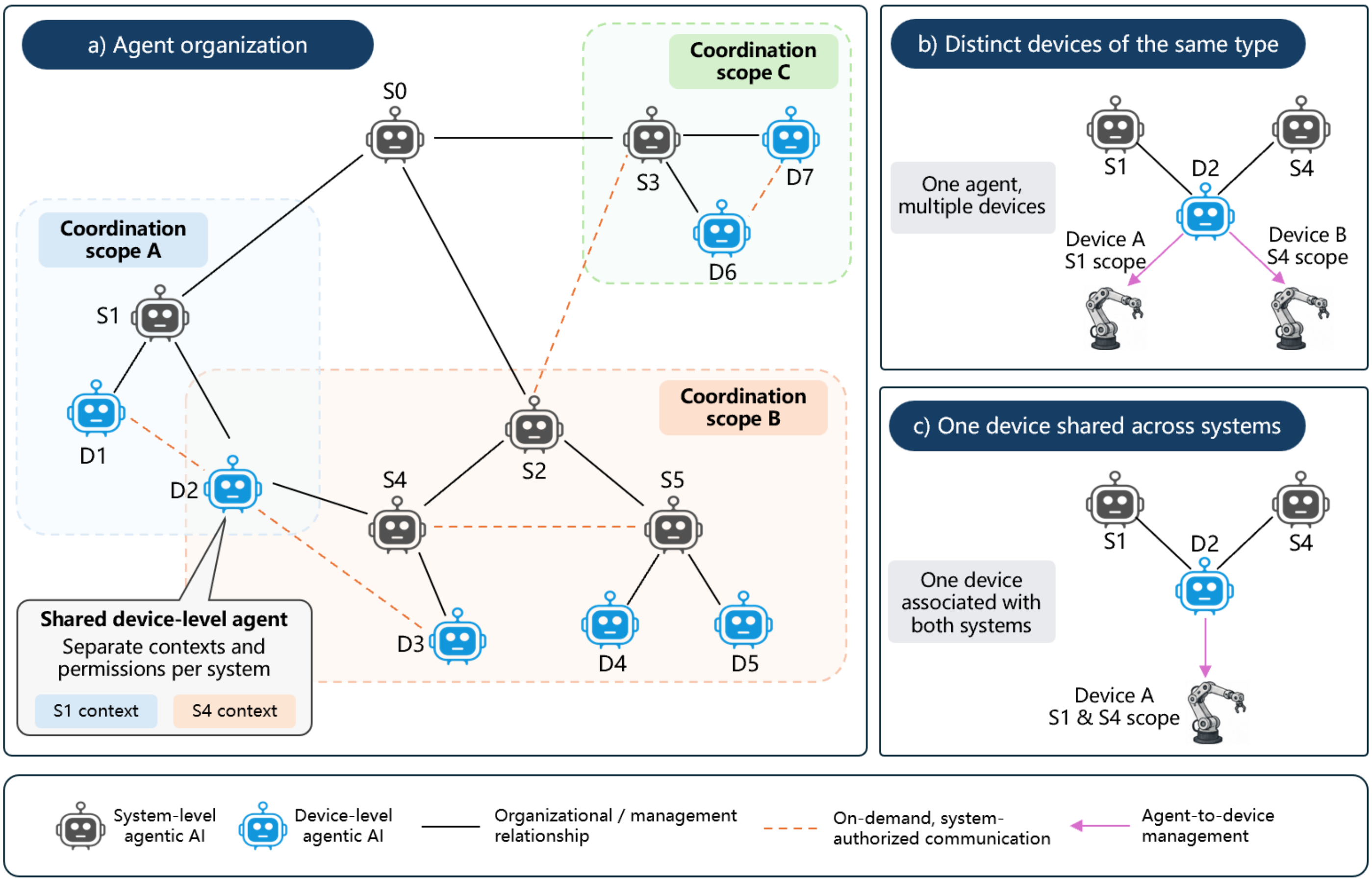

Figure.3 Organization of the network of agentic AIs: (a) agent network and coordination scopes;(b) one device-level agent managing distinct devices of the same type; (c) one device-level agent managing a device shared across systems.

A device-level agent may be associated with multiple system-level agents, with contexts and permissions isolated for each association. For example, D2 is associated with both S1 and S4. Agent associations also do not imply a one-to-one correspondence with physical devices. Fig.3b) shows D2 managing two distinct devices of the same type: Device A within S1's coordination scope and Device B within S4's scope. Each device maintains its own configuration and operating context.

Alternatively, Fig.3c) shows D2 managing a single physical device associated with both systems. Context isolation therefore separates system-specific interactions and authority without necessarily implying separate physical resources.

Moreover, the solid and dashed lines in Fig. 3(a) distinguish organizational relationships from on-demand communication links. Organizational relationships specify management and coordination responsibilities. Additional communication links enable direct collaboration when required and are authorized by the responsible system-level agents within configured permission boundaries. In particular, device-level agents can communicate directly when their system-level agents delegate the necessary permissions. Configuration thus governs the organizational structure and permissible interactions, while system-level decisions enable task-dependent communication.

Figure.4 presents a four-skill structure for both system-level and device-level agents in the AI-native ICS. The skills separate external interaction, foundational configuration, knowledge support, and operational implementation, with their scope determined by the agent's responsibilities.

Interaction skill provides the unified external boundary for users and other agents. It interprets requests, identifies targets and missing information, and selects and coordinates the appropriate internal skills. It also manages identity and capability exposure, authorized relationships, request correlation, and status and result reporting. Internal skill dependencies are handled directly, whereas cross-agent exchanges pass through this interaction boundary.

Configuration skill establishes and maintains the verified foundation required for operation. It acquires and checks resource inventories, execution environments, dependencies, communication interfaces, deployment bindings, and relevant spatial or temporal relationships, and performs permitted configuration or dependency installation. At the device level, it maintains instance-specific configurations and device-host bindings. At the system level, it maintains system composition, intercomponent relationships, and integration requirements. Configuration records preserve identity, ownership, provenance, and verification status, with affected evidence rechecked after changes.

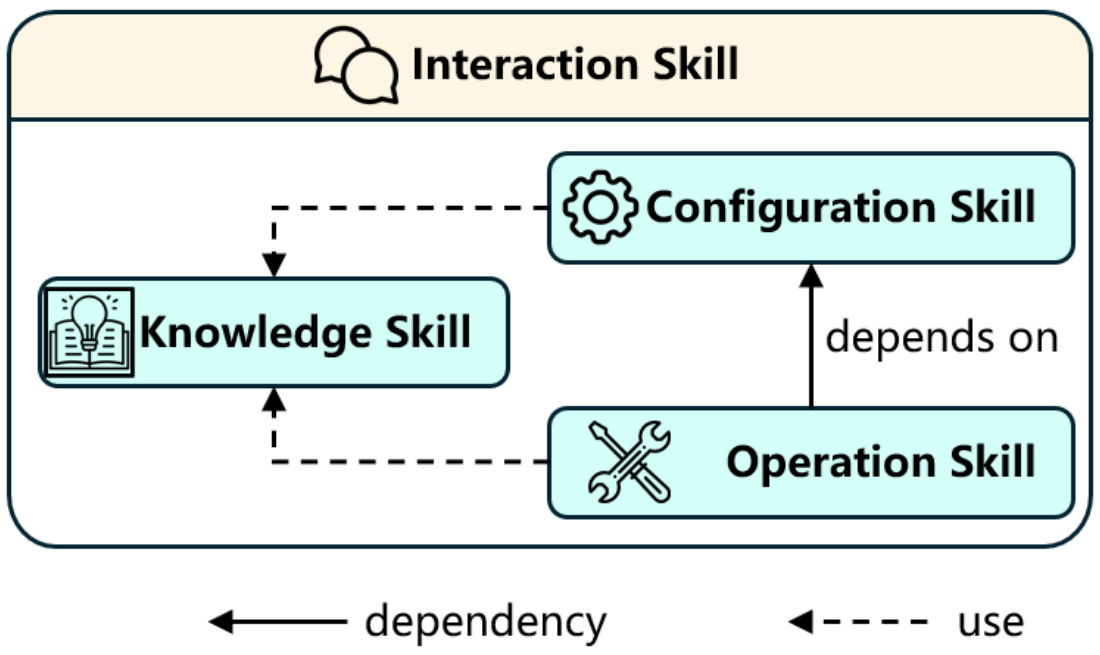

Figure.4 Four-skill structure of an agentic AI and the dependencies between skills

Knowledge skill supplies source-grounded information and, where available, verified resources. Device-level knowledge covers device properties, interfaces, constraints, and operating principles, drawing on authoritative sources such as official manuals, manufacturer-provided computational tools, and others. System-level knowledge covers control and supervisor methods, together with engineering knowledge relevant to system integration. The skill distinguishes explanatory information from executable capability and reports resource availability explicitly. It supports other skills as needed but does not itself configure environments, deploy applications, or operate physical devices.

Operation skill translates operational requirements into executable control implementation using verified configuration and available implementation resources. Device-level operation generates and verifies device-specific programs or parameterized modules. System-level operation constructs supervisory logic, coordinates module delivery, and integrates and verifies the complete control application. Within their respective scopes, both support authorized deployment and execution, maintain artifact revisions and execution evidence, and handle failures and recovery through defined interfaces.

As the solid arrow in Fig. 4 indicates, operation skill depends on configuration skill for verified resources, interfaces, and deployment conditions. The dashed arrows indicate that configuration and operation skills may use the knowledge skill when required. Interaction skill mediates external requests and exposes outcomes without replacing the verification responsibilities of the other skills.

## 2.3 Control System Construction, deployment and Evolution

As shown in Fig.5, control system construction begins by establishing the network of agentic AIs (Step 1). In light of task requirements, physical constraints, and other configuration information, the participating agents, their roles, and their organizational relationships are identified. The availability of these agents will be checked and the network covers the capabilities required for the task will be confirmed. This step establishes the task-specific control capability boundary by determining which capabilities the participating agents can provide and combine.

The confirmed network then interprets the task requirements and constraints and generates a candidate control application (Step 2). System-level agents construct the coordination logic and integrate the device-specific programs and interfaces supplied by device-level agents. The agents invoke build and verification tools to compile the application where required and check it against the applicable acceptance criteria (Step 3). Diagnostics and test results guide revisions to the candidate implementation. Once the applicable checks pass, the implementation is packaged as a versioned artifact.

Before deployment, the agents inspect the target execution environment and prepare the dependencies and device interfaces required by the selected revision (Step 4). In other words, this step establishes the task-specific execution environment boundary. Then the verified artifact is deployed to the prepared environment (Step 5), where the control application executes and governs the physical process through its control logic and device interfaces (Step 6).

AI-native control separates non-deterministic model reasoning during construction from deterministic supervisory logic during execution. The deployed application may still use AI for perception, estimation, or control. In that case, the supervisory logic specifies when AI outputs are accepted, how long the application waits for them, and what action it takes if they are invalid, delayed, or unavailable. These rules constrain how uncertain AI outputs affect physical execution.

Execution evidence, operational feedback, or changed requirements may initiate a new construction cycle. If a change affects the required capabilities or agent relationships, the network is reassessed. The agents revise the affected implementation or deployment configuration and repeat the relevant checks before deploying a new version.

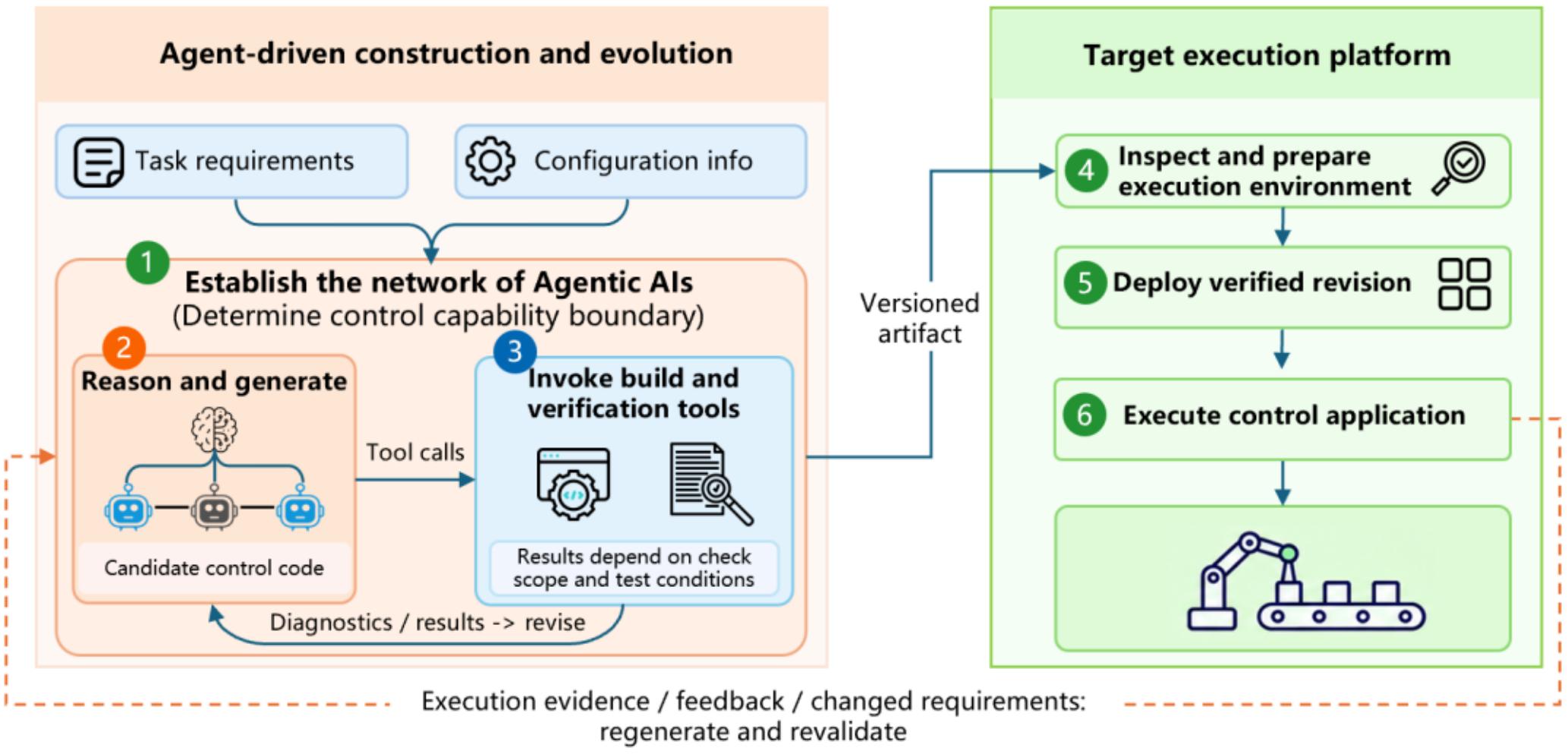

Figure 5 Control system construction, deployment, and evolution in AI-native ICS

# 3. Case study

## 3.1 Experimental Setup and Implementation

Figure 6 shows the experimental sorting workstation and its prepared configuration. The workstation comprises a Dobot Magician robot, a Hikrobot MV-CS050-10GC industrial camera, a motor-driven conveyor, two photoelectric sensors, and several storage locations. Sensor #1 detects materials at the camera inspection position, while sensor #2 detects their arrival at the robot pickup position.

A Debian 12 edge node equipped with an Intel Core i5-7300H CPU and 8GB DDR4 memory hosts a single OpenClaw instance in a Docker container. The agents use DeepSeek V4 Pro hosted on Pengcheng Cloud Brain for model inference. The Raspberry Pi 5 executes the integrated control application and is accessible from the edge node over SSH on the same local area network. The camera on the same local area network over GigE. The Raspberry Pi connects to the robot via USB and to a Modbus RTU I/O module via a USB-to-RS-485 adapter. The I/O module provides sensor inputs and conveyor control.

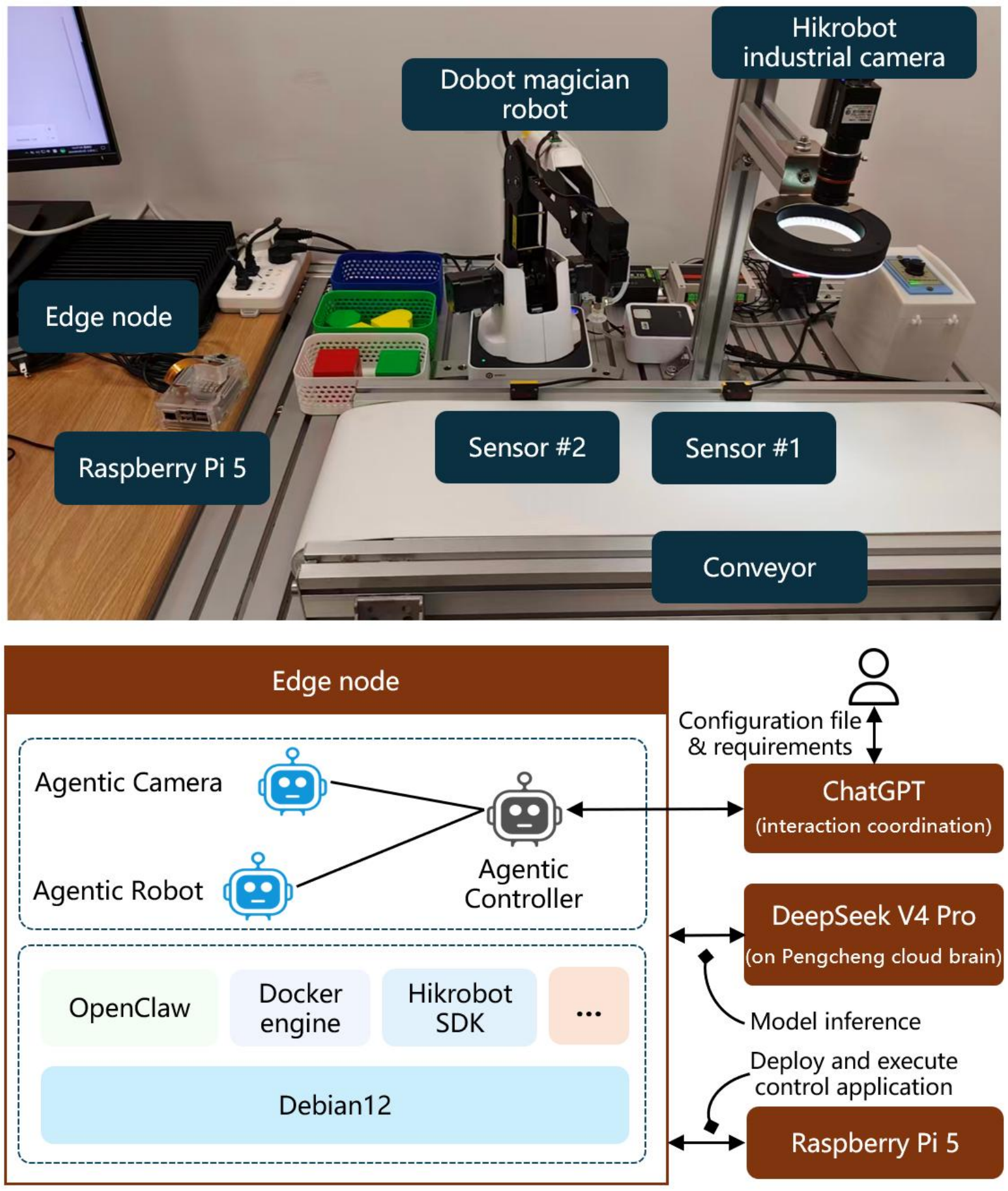

Figure 6 Sorting workstation and deployment of the agent and control environments

The implementation comprises three agents. Both the robot and the camera have their own device-level agents, Agentic Robot and Agentic Camera, respectively. They belong to a system-level agent: Agentic Controller. All three agents are built based on the four-skill structure and are hosted by the same OpenClaw instance, each with a dedicated workspace.

The operator provides clarification or on-site confirmation when needed. He or she supplies a configuration file, which contains specific task requirements, including material properties, storage coordinates and occupancy, sorting rules, and operating constraints, in addition to the experimental setup information mentioned above. In contrast to interacting with the Agentic Controller directly, ChatGPT is adopted to serve as the operator-facing interface. It understands and relays the configuration file to Agentic Controller, uses its response to formulate follow-up requests, and consults the operator when further input is needed. Agentic Controller coordinates its child device-level agents to generate, deploy, and execute control applications to meet certain task requirements.

### 3.2 Evaluation objectives and criteria

The case study serves two purposes. First, it validates that the AI-native ICS can be realized on a physical workstation through task-specific capability composition, execution environment preparation, and revision of the deployed control system. Second, it uses observations from construction, commissioning, and operation to identify where agent decisions require additional guidance or checks. These observations inform the analysis of implementation issues and research challenges in Section 4.

The evaluation uses evidence-based criteria for architectural feasibility. Each criterion specifies what must be demonstrated and which evidence supports that judgment. Configuration and interaction records establish how the agents organized and performed the work. Generated artifacts document the resulting control logic and interfaces. Logs and physical observations establish which implementations were executed and what behavior was observed.

- C1. Task-specific control capability. The participating agents and their contributions can be traced to the task requirements. The generated application combines the required device capabilities with task-specific coordination logic.
- C2. Task-specific execution environment preparation and deployment. Deployment records show how the agents inspected and prepared the prescribed execution environments or selected a suitable target where placement was left open. The deployed components use the required device and communication interfaces during operation.
- C3. Task-driven evolution. Changes in requirements can be traced to corresponding revisions in capability composition, control logic, interfaces, or deployment. The revised application exhibits the requested behavior while preserving the functions still required by the task.
- C4. Physical task completion. Task records and workstation observations show that the deployed application routes materials to the specified destinations or allows them to pass according to the sorting rule.

Two cases are launched. Case I establishes the baseline by constructing and deploying a sorting application without vision on a prescribed platform. Case II extends this baseline by integrating vision on a separate execution platform and revising the application to support attribute-based and selective sorting.

### 3.3 Case I: Construction of a sorting application without vision

Case I uses 50*50*50 mm materials of different colors. The sorting rule assigns them in sequence to the first three of the four configured storage locations, regardless of color. The task therefore requires conveyor coordination and robot pickup and placement, with no need for visual recognition. Agentic Controller and Agentic Robot form the agent network for this case and jointly establish its control capability boundary.

As shown in Fig.7a), the configuration file prescribes the Raspberry Pi 5 as the execution target. The two agents establish the necessary execution environment on that platform by checking the available interfaces and preparing the required dependencies. Then, they generate task-driven control applications. According to the logs, specifically, Agentic Robot installs the task-required libraries and supplies the robot position files, a parameterized motion program, and its callable interface through its configuration and operation skills. Agentic Controller generates the sorting logic and integrates the robot interface with Modbus RTU sensor acquisition and conveyor control after inspecting the status of sensors and conveyor.

Figure 7(b) shows the finite-state machine reconstructed from the generated sorting controller. When DI2 (Sensor#2) detects a material at the pickup position, the controller requests a conveyor stop and allows the material to settle before initiating sorting. It evaluates candidate storage locations and calls the robot synchronously for an eligible destination. After a successful placement, it advances the starting point for the next destination search and waits for DI2 to clear before requesting a conveyor restart. If no candidate can accept the material, the controller enters FullStorage and holds the conveyor stopped pending operator intervention. During commissioning, the operator observed a storage-capacity rejection followed by a conveyor stop and manual restart after the storage condition was resolved.

The video frames and task records in Fig. 7(c) document completed placements at the specified destinations. Together with the generated implementation and logs, they show how the two agents prepared the prescribed environment and combined device functions with task-specific coordination logic to execute the sorting application on the Raspberry Pi 5.

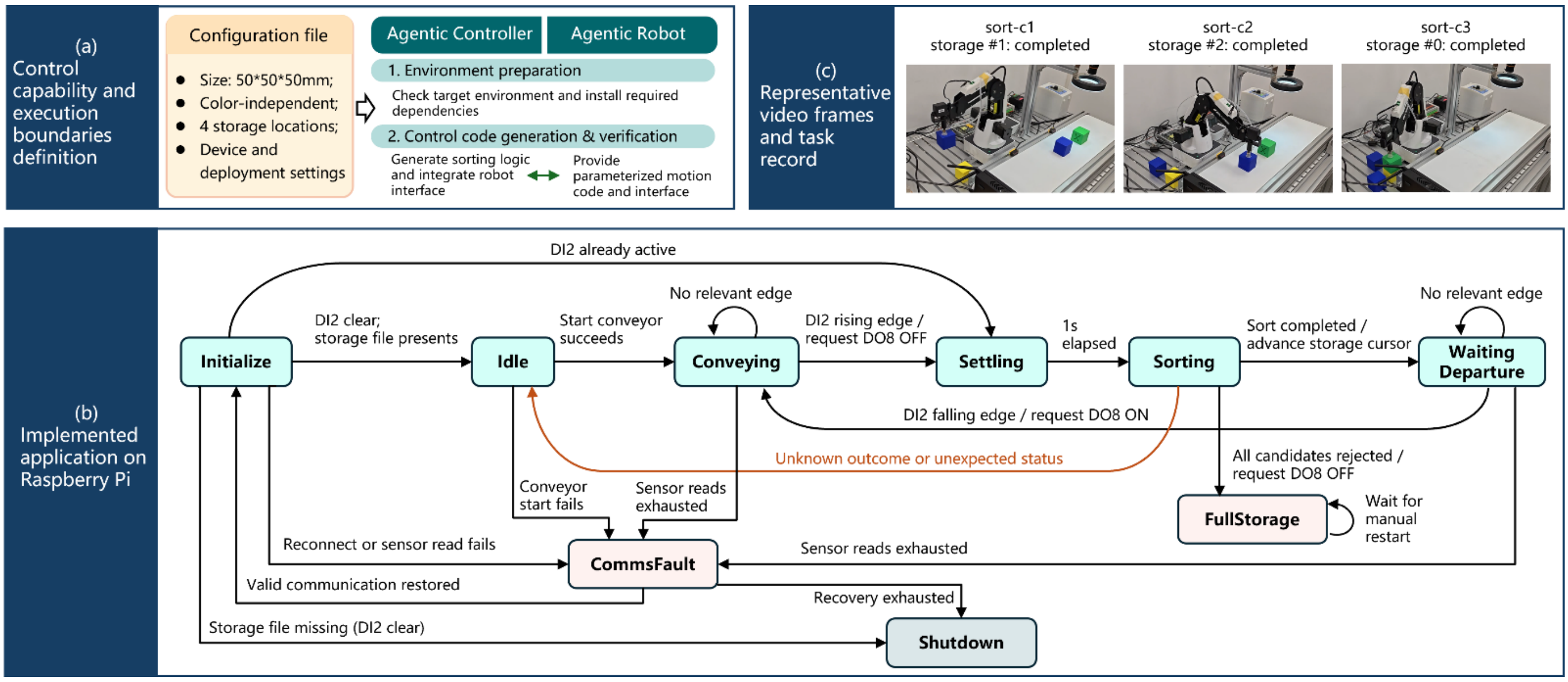

Figure 7 Case I boundaries definition, generated controller states, and representative execution evidence

## 3.4 Case II: Vision integration and task-driven evolution

Case II makes sorting dependent on material attributes. This change extends the control capability boundary to include visual inspection and brings Agentic Camera into the network alongside Agentic Controller and Agentic Robot. The Raspberry Pi 5 remains the prescribed target for the sorting controller and robot program. The configuration leaves the placement of the inspection program open. Agentic Controller and Agentic Camera coordinate its deployment on the edge node, where Agentic Camera prepares the Hikrobot SDK and exposes an interface that acquires and analyzes five frames per item (Fig. 8(a)). The controller on the Raspberry Pi 5 invokes the inspection program over SSH and reuses the robot interface for pickup and placement.

Compared with the Case I controller in Fig. 7(b), the control flow in Fig. 8(b) adds inspection at DI1 (Sensor#1) and associates the resulting attributes with the arrival at DI2. At the inspection position, the controller requests a conveyor stop, creates a material record, and associates the recognition result with that record before conditionally resuming the conveyor. At the pickup position, it matches the arrival to the oldest queued record and applies the sorting rule. The generated coordination logic thus connects an upstream observation to a downstream control decision across the two execution environments.

In Task I, the application sorts all materials according to either shape or color. Changing the selected attribute changes destination selection while retaining the pickup sequence. Task II changes the conditions for robot dispatch. In the demonstrated trial, cubes follow the sorting path and non-cubes continue downstream without robot pickup. This revision retains the inspection and robot interfaces while changing how the controller uses the recognition result. Figure 8(c) shows representative observations of color-based sorting, cube pickup, and non-cube passage.

The generated implementations also differ in robot scheduling and storage checking. In Task I, the controller dispatches the robot call to a background thread, allowing its main loop to continue polling DI1 and DI2 and checking for the robot result. In Task II, the controller calls the robot synchronously for selected materials, then records the result and checks whether DI2 has cleared. Case I represents exhausted storage explicitly as FullStorage. In Case II, the robot module checks whether the selected destination can accept the material during trajectory planning and returns a rejection if placement is infeasible. These differences concern how the generated implementations allocate responsibilities and coordinate execution. Section 4.2 discusses the consistency and traceability of these design decisions.

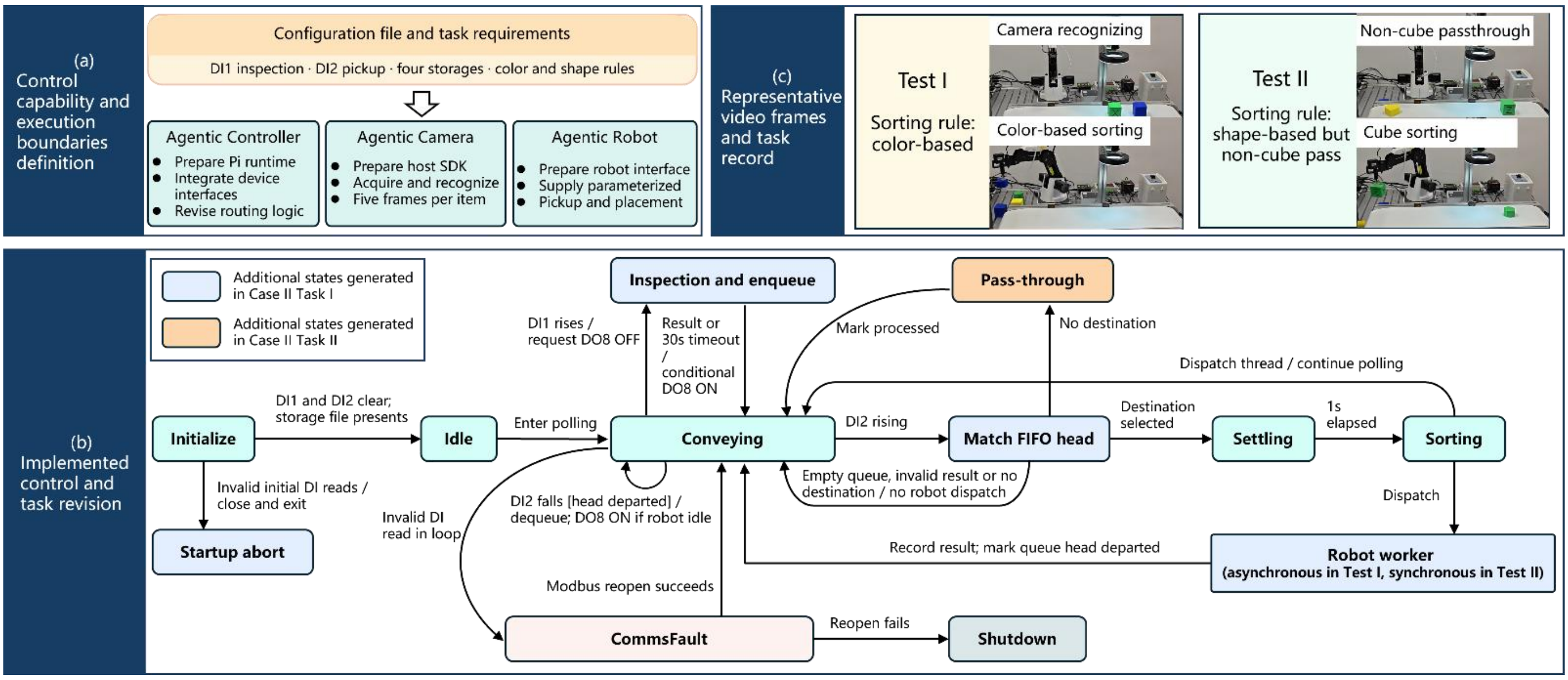

Figure 8 Case II boundaries definition, generated controller states, and representative execution evidence

Overview, Table I relates the evidence from both cases to the criteria defined in Section 3.2.

Table 1. Evidence against the evaluation criteria in Cases I and II

| Evaluation criterion | Case I evidence | Case II evidence | Scope of evidence |
|---|---|---|---|
| **C1. Task-specific control capability** | Agentic Controller generated sorting logic and integrated the motion interface supplied by Agentic Robot (Fig. 7(a–b)). | Agentic Camera supplied the inspection interface. Agentic Controller linked recognition results to routing and robot dispatch while reusing the robot interface (Fig. 8(a–b)). | Shows how the agent network and generated coordination logic realize different capability boundaries on the same workstation. |
| **C2. Task-specific execution environment preparation and deployment** | The agents prepared the prescribed Raspberry Pi 5 environment. The application used Modbus RTU for sensor and conveyor I/O and invoked the robot interface during sorting. | The agents kept sorting and robot execution on the Raspberry Pi 5 and placed inspection on the edge node. The deployed controller received recognition results over SSH. | Demonstrates preparation of a prescribed target and placement of a function whose execution target was left open by the configuration. |
| **C3. Task-driven evolution** | Provides the baseline application without vision. | Integrates vision, changes the attribute used for | Establishes changes in capability composition and |
| | | routing, and revises dispatch logic to permit selective pickup and passage (Fig. 8(b–c)). | application behavior in response to revised requirements. |
| **C4. Physical task completion** | Task records and representative video frames document completed placements (Fig. 7(c)). | Trial records and representative video frames document sorting of selected materials and downstream passage of excluded materials (Fig. 8(c)). | Confirms physical behavior for the demonstrated sorting rules and operating conditions. |

# 4. Discussion and challenges

## 4.1 From Open Control to AI-Native Industrial Control

Over the past four decades, the evolution of ICS is closely linked to advances in computing. Fig.9 summarizes three overlapping development directions: open control, networked control, and virtual control.

Open control primarily aims to make control capability easier to modify. Most studies use modular software and standardized interfaces to support the addition, replacement, and recombination of control modules [12-15]. However, a persistent difficulty is that modules with compatible interfaces may make different assumptions about execution. Completion criteria, state dependencies, and fault responses can affect their integration even when command formats agree. Changes to a module may therefore require corresponding changes to system coordination and renewed validation of the resulting behavior.

Networked control extends deployment flexibility by separating control computation from the equipment it serves. Distributed control functions can use edge or cloud resources, while shared computing and communication infrastructure can support multiple control tasks [17, 18]. The main constraint is the effect of communication delay, jitter, and packet loss on closed-loop behavior [16]. Cloud-based CNC addresses this constraint by retaining time-critical servo functions locally while assigning selected machining functions to remote resources [17]. Control-as-a-service approaches coordinate task execution with network traffic to meet timing requirements [18]. In both cases, a deployment is feasible only if the complete path from sensing to actuation meets the requirements of the physical process.

Virtual control aims to improve resource utilization and simplify controller management by allowing multiple runtimes to share a hardware platform. Virtual machines or containers package execution environments, and orchestration mechanisms manage their allocation and deployment. Commercial virtual PLCs and research on containerized industrial applications illustrate this approach [19-22]. Packaging improves portability, but the suitability of a host still depends on device access, scheduling, and interference between workloads. Moving a packaged controller can therefore require renewed assessment of its execution timing even when its control code is unchanged.

Cloud–fog automation illustrates the convergence of networked and virtual control. It combines virtualized computing with industrial networks and treats the placement of control functions, the allocation of computing resources, and the provision of communication services as related design decisions [23,24]. Our earlier work on platform-enabled CNC makes the connection to open control explicit. The platform manages CNC function modules, computing resources, and machine tools separately, allowing a controller to be composed for a machining task, deployed on suitable resources, and assigned to the corresponding machine tool[25].

As Fig. 9 illustrates, AI-native ICS builds on this convergence by bringing the definition of control capability and execution environment boundaries within the scope of AI reasoning and action. Given task requirements and a configuration of available resources, agents determine which control capabilities are required, construct the application that coordinates them, and prepare a suitable execution environment. When requirements change, the agents can revisit these decisions together. This extends architectural flexibility to the process by which the control system itself is constructed and revised.

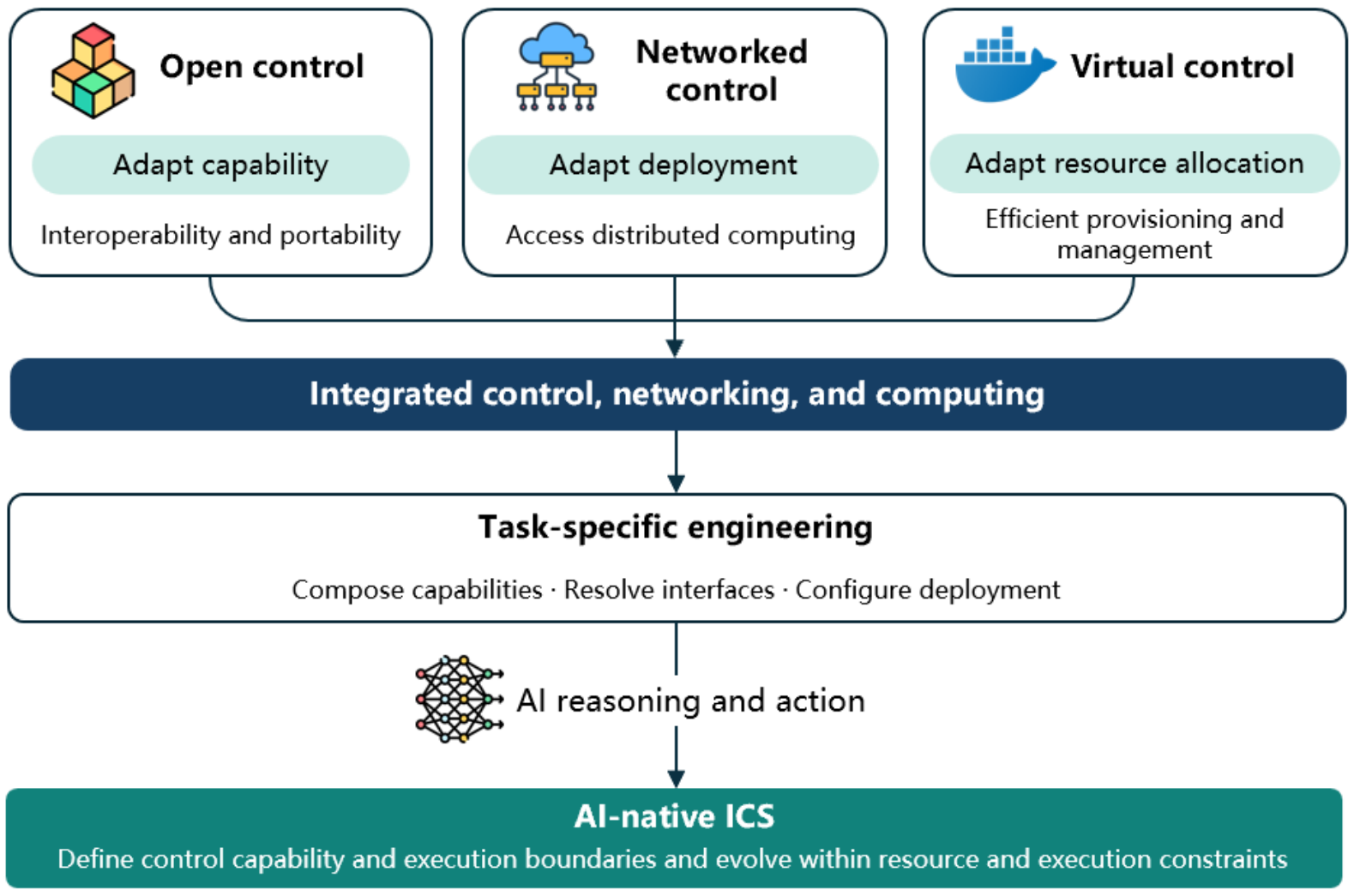

Figure 9. From open, networked, and virtual control to task-specific control system construction and evolution by AI agents.

At the functional level, task requirements determine which agents need to participate and how they must be associated. Confirming their roles, relationships, and availability establishes the network of agentic AIs for the task. The capabilities that these agents can provide and combine define the control capability boundary. Within this boundary, different control applications are generated based on certain requirements. For example, control applications for Task I and Task II in Case II are both generated within the control capability scope made up of Agentic controller, Agentic Robot, and Agentic Camera. A change in control logic therefore requires a change in the agent network only when the revised task calls for capabilities or relationships beyond those already available.

At the deployment level, task requirements determine what the execution environment must provide, including device access, software dependencies, and timing behavior. The same considerations govern whether a control component should run natively or in a container. Agents establish the execution environment boundary by relating these requirements to the resources and permissions specified in the configuration. When a target is prescribed, they inspect and prepare it for the required functions. When placement is open, they also select a suitable target and deployment arrangement. The resulting boundary defines where and under what conditions the control application can execute.

These two boundaries can evolve together, with the task change determining the scope of revision. Adding vision extended the agent network, introduced an execution environment on the edge node, and changed how the controller tracked materials. The subsequent selective sorting task reused the same hosts and device interfaces but changed when the conveyor stopped and when the robot was dispatched. In the process presented in Section 2.3, agents use the revised requirements and execution feedback to identify the affected functions and deployment conditions, then revalidate the implementation before deploying the revision.

### 4.2 Implementation Issues Revealed by the Case Study

The physical trials establish that agents can construct and revise a working sorting application. The commissioning records also reveal where decisions about control logic, delegation, diagnosis, and verification required additional guidance.

The agents did not consistently explain why control design decisions changed across applications. In Case II, Task I dispatched robot operations asynchronously, whereas Task II used a synchronous call, as Case I does. This choice determines whether the main loop can continue polling sensors during robot motion. Responsibility for storage handling also changed: the Case I controller entered `FullStorage` when all candidate locations were rejected, while the Case II robot module checked the selected location during trajectory planning and could reject the placement. Figures 7(b) and 8(b) document these differences, but the records do not establish which task requirement or execution constraint motivated them. Task-driven evolution therefore calls for methods that trace design changes to their rationale and identify the behavior that should be preserved across revisions.

The configured agent network did not consistently govern who acted or what information was available to the responsible agent. In Case I, Agentic Controller used libraries installed by Agentic Robot to write motion code itself, although its instructions assigned robot operations to Agentic Robot. Delegation occurred only after an explicit reminder. In Case II, Agentic Robot initially lacked the device identity specified in the system configuration and selected a local identifier until a later handoff resolved the discrepancy. These observations motivate research on how to enforce delegated authority at the point of action and transfer the corresponding device identity, permissions, and configuration when a task is assigned.

The agents could repair implementation defects, but sometimes required guidance to identify the decisive diagnostic check. In Case I, Agentic Controller initially suspected incorrect parity or slave addressing after receiving Modbus frames that failed CRC validation. An external request prompted it to inspect the transmitted and received bytes. It then found that the receive routine read only five bytes of a six-byte response and corrected the defect. In Case II, the agents traced discarded recognition results to a status-value mismatch and verified the correction using saved images. Further research should examine how agents assess the reliability of observations and select tests that distinguish among plausible causes before revising an implementation.

Some verification claims extended beyond the property actually checked. In Case II, a deployment report inferred that blocking the main loop during image recognition would preserve sensor-event detection because the controller issued a conveyor stop request. Source inspection established the order of the commands. The conclusion about event detection also depended on when the conveyor physically stopped and whether a sensor could change state before polling resumed. This case motivates verification methods that link each behavioral claim to its software and physical assumptions and use timing analysis or execution observations to check those assumptions.

## 4.3 Research Challenges for AI-native ICS

Based on the overview of AI-native ICS in Section 2 and the issues discussed in Section 4.2, the following challenges are summarized.

Translating task requirements into control behavior specifications. A task description may specify the intended production outcome while leaving completion semantics, timing requirements, fault responses implicit, and other information. For example, a sorting rule describes which materials to pick while leaving open whether sensor monitoring must continue during robot motion and how the controller should respond to a rejected placement. NL2TL and nl2spec investigate the translation of natural-language requirements into temporal specifications and the clarification of ambiguous interpretations [26,27]. ACIES further decomposes system requirements into verifiable component contracts[28]. For AI-native ICS, the challenge is to distinguish required behavior from implementation choices. If sensor monitoring must continue during motion, the agent should record that obligation and explain how its scheduling choice satisfies it. If the requirement leaves the choice open, the agent should identify the assumption behind its decision. This distinction makes it possible to determine which behavior must be preserved and which design choices may change when an application is revised.

Verifying system behavior under physical and timing constraints. A verification claim must identify the property checked and the conditions under which it holds. LLM4PLC integrates external verification into control program generation[8], while contract-based synthesis examines whether component guarantees satisfy a system requirement[28]. The inspection example in Section 4.2 illustrates a further obligation: establishing that the program issues a stop request is a source-level check, whereas establishing that no sensor event is missed also requires evidence about stopping time, event duration, and input acquisition. Research is needed on how agents identify these dependencies, select an appropriate combination of analysis and physical checks, and determine which evidence remains valid after a revision. Where AI outputs participate in the control loop, runtime assurance offers a means of enforcing operating constraints [29]. The response deadline and fallback action must then be justified against the dynamics of the controlled process.

Enforcing delegated authority and maintaining shared configuration. Establishing the network of agentic AIs defines responsibility, but each handoff must also convey the device identity, configuration, and permissions needed to exercise it. Multi-agent frameworks support programmable interaction [30], while studies of agent failures identify persistent difficulties in coordination and verification [31]. Runtime policies and controls on tool use provide mechanisms for enforcing permitted actions [32,33,37]. Industrial control adds a physical dimension to delegation. In Fig. 3(c), two system-level agents may each be authorized to use the same device while their requested operations are incompatible with its current state. Research must connect communication permissions with authority at the point of device access and define how that authority is granted, transferred, or revoked while an operation is in progress.

Preparing execution environments that satisfy control requirements. A host with the required SDK and sufficient computing capacity may still be unsuitable if communication delay, scheduling interference, or loss of device access prevents a timely response. Control-as-a-service, cloud–fog automation, and industrial fog platforms address complementary aspects of computation, networking, and resource provision [18,23,24,38]. AI-native ICS requires agents to derive placement criteria from the behavior specification, inspect candidate targets, and validate the selected arrangement over the complete sensing-to-actuation path. When a component moves or a workload is added to a shared host, the agents must identify which timing and availability assumptions require revalidation. This applies both to a target prescribed by the configuration and to one selected by the agents.

Diagnosing faults and revising a running control system. Feedback-based methods use previous observations or verification results to guide repair [28,35], and distributed tracing can relate observed execution to specified expectations [39]. The Modbus incident mentioned in Section 4.2 shows why an agent must also choose an observation that distinguishes among plausible causes. Inspecting the received bytes resolved an ambiguity that repeated reports of CRC failure could not. For industrial deployment, diagnosis must further establish which application revision ran and what physical action occurred. If a robot command loses its acknowledgement, the device may already have completed the motion. Before retrying or deploying a correction, the agent must reconcile the command record with the observed device state. Research on controlled evolution should therefore connect diagnostic reasoning, revision history, and physical recovery within the same workflow.

Developing the AI-native infrastructure described in Section 2.1 is a further challenge. As agent runtimes, harnesses, and model services evolve, the infrastructure must continue to support reliable coordination among agents and controlled access to the engineering tools they use to construct and deploy control applications.

## 5. Conclusion and future work

This paper proposed an AI-native ICS in which task requirements determine the participating agents and their associations.

Within the resulting capability boundary, agents compose device functions, generate coordination logic, and prepare the resources on which the control application executes. Agent-driven construction and application execution serve distinct roles, allowing implementations to evolve while physical operations remain governed by the deployed logic.

The sorting workstation demonstrated both a change in capability and a change within an existing capability boundary. Introducing vision added a camera agent, an inspection program on the edge node, and logic linking inspection to pickup. Selective sorting then changed conveyor and robot dispatch decisions while retaining the same agents and execution platforms. The evidence establishes architectural feasibility for these tasks and deployment conditions.

Future work will link generated design decisions to explicit requirements for completion, timing, and failure handling and investigate how to verify those requirements across software and physical execution. A further priority is to enforce delegated permissions and maintain authoritative device configurations when agents share equipment. Experiments across additional workstations and deployment targets will examine which guarantees remain valid after changes to the application, its execution environment, or the agent infrastructure.